\documentclass[apj]{emulateapj_nt}
\slugcomment{December 2017} 

\usepackage{dcolumn}
\usepackage{bm}
\usepackage{graphicx}
\usepackage{amssymb,amsmath}
\usepackage{epsfig}
\usepackage{color}
\usepackage{wasysym}
\usepackage{natbib}
\usepackage{twoopt}
\usepackage{dashrule}
\definecolor{slateblue}{rgb}{0.1,0.22,0.58}
\usepackage[breaklinks=true,colorlinks=true,linkcolor=slateblue,citecolor=slateblue,urlcolor=slateblue,pdfauthor={Tomassetti},pdftitle={AbarSnr}]{hyperref}
\def\MyTitle#1{{\emph{#1} --- }} 
\definecolor{ColorTitle}{cmyk}{0,.88,.77,.40}
\newcommand{\etal}{et al.}
 
\newcommand{\ie}{\textit{i.e.}} 
\newcommand{\pbarp}{\ensuremath{\bar{p}}/\ensuremath{p}}
\newcommand{\BC}{\textrm{B}/\textrm{C}} 
\newcommand{\Li}{\textrm{Li}} 
\newcommand{\Be}{\textrm{Be}} 
\newcommand{\B}{\textrm{B}} 

\makeatletter
\def\blfootnote{\gdef\@thefnmark{}\@footnotetext}
\makeatother

\begin{document}
\title{Fresh Insights on Cosmic-Ray Propagation from the new AMS Data}
\author{Nicola Tomassetti\,$^{1}$, Jie Feng\,$^{2,\,a}$, Alberto Oliva\,$^{3}$}  
\address{$^{1}$\, Department of Physics and Earths Science, Universit{\`a} di Perugia, and INFN-Perugia, 06100 Perugia, Italy;}
\address{$^{2}$\, School of Physics, Sun Yat-Sen University, Guangzhou 510275, China;} 
\address{$^{3}$\, Centro de Investigaciones Energ{\'e}ticas, Medioambientales y Tecnol{\'o}gicas CIEMAT, 28040 Madrid, Spain}
%%%%%%%%%%%%%%%%%%%%%%%%%%%%%%%%%%%%%%%%%%%%%%%%%%%%%%%%%%%%%%%%%%%%%%%%%%%%%%%%%%%%%%%%%%%%%%

%%%%%%%%%%%%%%%%%%%%%%%%%%%%%%%%%%%%%%%%%%%%%%%%%%%%%%%%%%%%%%%%%%%%%%%%%%%%%%%%%%%%%%%%%%%%%%%%%%%%%%%%%%
\maketitle
\begin{figure*}[!t]
\epsscale{1.16}
\plotone{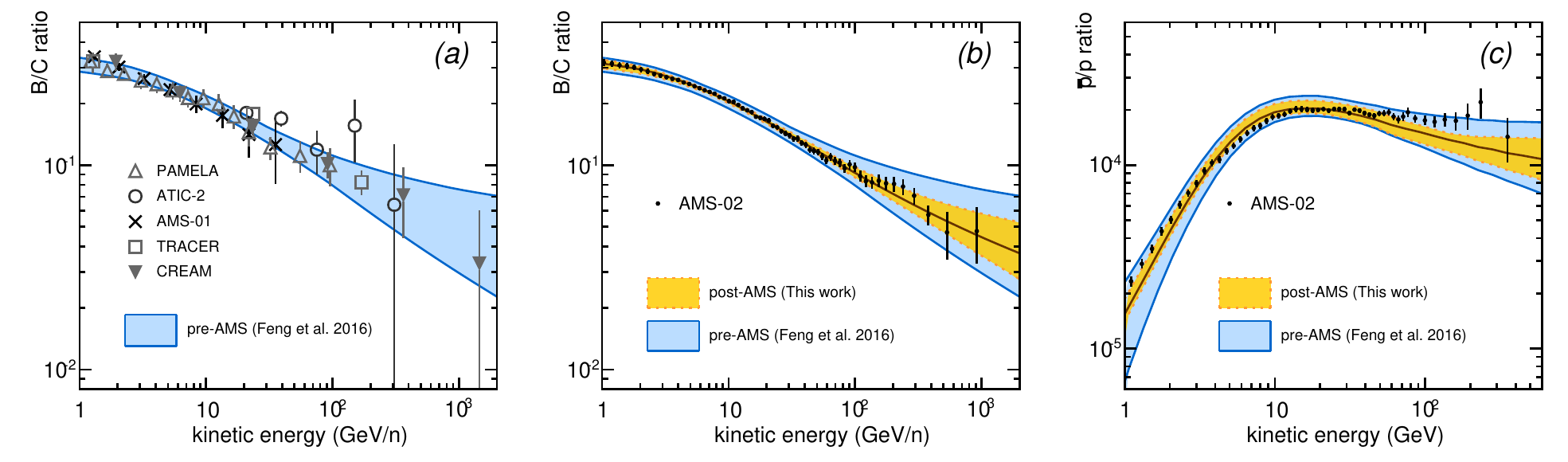}
\caption{\footnotesize%
  Calculations for the \BC{} (a, b) and \pbarp{} (c) ratios in comparison with the AMS data. 
  The blue shaded bands are the uncertainties obtained from our pre-AMS data (a) analysis \citep{Feng2016}. 
  The orange band is from this work, \ie, using the AMS data only.}
\label{Fig::ccSecPriRatioVSEkn}
\end{figure*}

%%%%%%%%%%%%%%%%%%%%%%%%%%%%%%%%%%%%%%%%%%%%%%%%%%%%%%%%%%%%%%%%%%%%%%%%%%%%%%%%%%%%%%%%%%%%%%%%%%%%%%%%%%
The Alpha Magnetic Spectrometer (AMS) experiment has released new measurements of primary and
secondary nuclei in cosmic rays (CRs) at the TeV energy scale \citep{Aguilar2016BC,Aguilar2017CO}.
Using these data, we present new results based on a global Bayesian analysis
of our \emph{two halo} model of CR transport \citep{Feng2016}.
\\[0.18cm]
%%%%%%%%%%%%%%%%%%%%%%%%%
\MyTitle{Background}  %%%
%%%%%%%%%%%%%%%%%%%%%%%%%
%
Measurements of energy spectra of \emph{primary} (\ie, accelerated in Galactic sources) 
and \emph{secondary} (\ie, spallogenic) nuclei are essential to investigate
the acceleration and transport properties of CRs in the Galaxy, 
thereby providing the \emph{astrophysical background} for the search of dark matter via cosmic antimatter.
In \citet{Feng2016}, we made use of a large compilation of \emph{pre-AMS} data on the \BC{} ratio
to constrain the parameters of a \emph{two halo model} of CR propagation.
This model explains the CR spectra in terms of two propagation regions having different energy-dependent diffusion coefficients:
$D_{i}\propto\,E^{\delta}$ for the inner region near the Galactic disk, where the turbulence is generated by SN explosions,
and $D_{o}\propto\,E^{\delta+\Delta}$ (with $\Delta\approx\,0.55$) for the outer region away from the disk, 
where the turbulence is driven by CRs \citep{Tomassetti2015TwoHalo}.
Our previous results suggested a surprisingly shallow diffusion for CRs in the inner region 
($\delta=0.18\pm\,0.13$ within $l\approx$\,900\,pc) but, owing to uncertainties in the \BC{} data, 
these results were still consistent with the conventional Kolmogorov-like expectation $\delta=1/3$. 
In this Note, we present fresh results based on an updated Markov-Chain Monte-Carlo sampling
of new \BC{} data from AMS \citep{Aguilar2016BC}.
\\[0.18cm]
%%%%%%%%%%%%%%%%%%%%%%%%%%
\MyTitle{New results}  %%%
%%%%%%%%%%%%%%%%%%%%%%%%%%
%
The AMS data lead to narrow probability density functions for the CR transport parameters.
The resulting uncertainty band for the \BC{} ratio is shown in Fig.\,\ref{Fig::ccSecPriRatioVSEkn}b
in comparison with the one estimated using pre-AMS data of Fig.\,\ref{Fig::ccSecPriRatioVSEkn}a.
The posterior mean of the near-disk diffusion exponent is $\delta=0.18\pm\,0.05$. 
While this agrees with our earlier findings, the value $\delta\equiv\,1/3$ is now excluded at 95\,\%\,CL. 
Taken at face value, our results disagree with the
observations made by \emph{Voyager-1}  in the insterstellar medium, 
which reported a Kolmogorov-like spectrum of magnetic turbulence \citep{Burlaga2015}.
This tension can be relieved by the incorporation of a hard \emph{source component} of \Li-\Be-\B{} nuclei,
generated by nuclear interactions of CRs during acceleration \citep{BlasiSerpico2009,MertschSarkar2014}.
The production rate is regulated by the plasma density upstream the CR sources, $n_{1}$, which is not well known.
Thus, have included interactions in our CR acceleration calculations.
We have repeated the fits after fixing $\delta\equiv\,1/3$ and leaving $n_{1}$ as free parameter. 
Interestingly, the results are found to be consistent with the average matter density of the Galaxy,
$n_{1}\sim$\,1\,cm$^{-3}$ for CR sources of typical age $\tau_{\rm SN}\sim$\,20-30\,kyr,
as one expects if the CR flux is provided by a large ensemble of supernova remnants \citep{TomassettiOliva2017}.
Similar findings were reported in \citet{Aloisio2015}, where a
model of nonlinear CR propagation was tested against the preliminary AMS data.

In Fig.\,\ref{Fig::ccSecPriRatioVSEkn}c, we present the \pbarp{} astrophysical background calculated with the new \BC-driven constraints.
In comparison with the pre-AMS \BC-driven results, we found an increased discrepancy:
the flat \pbarp{} behaviour reported by AMS cannot be recovered by our model.
We also note that, as found in our past works,
the inclusion of primary antiprotons (from proton-gas interactions inside sources) does not resolve this tension 
satisfactorily \citep{TomassettiOliva2017,TomassettiFeng2017}.
The uncertainties in the \pbarp{} calculations are now dominated by cross-section 
data rather than by CR data, and, at low energy, by solar modulation. 
While these uncertainties could recover the tension,  there is yet some room left to squeeze in a potential dark matter component. 
%
%%%%%%%%%%%%%%%%%%%%%%%%%%%%%%%%%%%%%%%%%%%%%%%%%%%%%%%%%%
%%%%%%%%%%%%%%%%%%%%%%%%%%%%%%%%%%%%%%%%%%%%%%%%%%%%%%%%%%
\\[0.18cm]
%%%%%%%%%%%%%%%%%%%%%%%%%%
\MyTitle{Conclusions}  %%%
%%%%%%%%%%%%%%%%%%%%%%%%%%
%
The new data released by AMS can be described well by a two-halo model of CR propagation,
but they point to a very shallow diffusion, for CRs near the disk,
which disagrees with the \emph{Voyager-1} observations of interstellar turbulence. 
This tension can be relieved if secondary CR production inside sources is included.
With the improved constraints, we have also found that a tension 
between \pbarp{} data and astrophysical background calculations starts to emerge. 
Because cross sections are dominating the uncertainties, LHC measurements on
antiproton and antineutron production will very precious for clarifying the situation.
\\[0.18cm]
{\footnotesize%
%N.T. acknowledges the European Commission for support under the H2020-MSCA-IF-2015 action, grant No.707543-MAtISSE.
A.O. acknowledges CIEMAT, CDTI and SEIDI MINECO under grants ESP2015-71662-C2-(1-P) and MDM-2015-0509.
NT acknowledges support from the MAtISSE project.
This project has received funding from the European Union's Horizon 2020 research and innovation programme under the Marie Sklodowska-Curie grant agreement No 707543.
}

%%%%%%%%%%%%%%%%%%%%%%%%%%%%%%%

\end{document}